\documentstyle[prl,aps,epsf,two column]{revtex}

\include{psfig}

\title{
Quantum Spin Fluctuations as a Source of Long-Range Proximity
Effects in Diffusive Ferromagnet-Superconductor Structures}  
\author{A. Kadigrobov$^{1,2}$, R. I. Shekhter$^{1}$, and M.~Jonson$^{1}$,\\
}
\address{
 $^{1}$Department of Applied Physics,
 Chalmers University of Technology and G\"oteborg University, 
SE-412 96 G\"oteborg, Sweden\\
 $^{2}$B. I. Verkin Institute for Low Temperature Physics \& 
Engineering, \\
National Academy of Science of Ukraine, 47 Lenin Ave., 310164 Kharkov, 
Ukraine\\
}

\begin{document}

\date{22 December 2000}
\maketitle

\begin{abstract}
We  show that  quantum spin fluctuations in inhomogeneous ferromagnets 
drastically affect the Andreev reflection of electrons and holes at a 
ferromagnet-superconductor interface.
As a result a strong long-range proximity effect appears, 
associated with electron-hole spin triplet correlations and 
persisting on a lenght scale 
typical for non-magnetic materials, but anomalously large 
for ferromagnets. 

\end{abstract}

\vspace{3mm}
 
\noindent
In recent years much attention has been paid to normal 
conductor-superconductor (N/S) structures  in which transport properties
of the normal conductor are much modified in the vicinity of 
the superconductor  (for a review,  see paper  \cite{Lambert}).
The origin of this "proximity effect" has to do with  
 correlations of normal-metal electrons and holes caused by 
Andreev reflection at the inteface with the superconductor. 
An important feature of such Andreev scattering, which converts 
electrons into holes and vice versa, is the selection rule that requires 
 the energies of the electron and hole (as measured from the Fermi 
energy) - as well as  their spin projections -  to be equal in 
magnitude but  opposite in sign. While the spin selection rule 
is irrelevant for the proximity effect in non-magnetic normal 
materials, the energy selection  results in a destructive 
interference between the electron and hole states  corresponding 
to a decay of the electron-hole correlations at distances of 
order $L_{T} = \sqrt{\hbar D/kT}$ from the N/S interface ($D$ 
is the diffusion constant 
of the normal conductor, $T$ is temperature).

In ferromagnetic materials electrons and holes acquire an 
additional exchange energy, which is sensitive to the direction 
of the spin.  Hence the spin selection rule for Andreev reflection
becomes relevant. As a result  the proximity effect decays on the 
much smaller  length scale $L_{I_0} = \sqrt{\hbar D/ I_0}$  
($I_0$ is the exchange energy). Typically,  $I_0$ exceeds
$k_BT$ by several orders of magnitude, resulting in  a drastic 
reduction of the proximity effect in magnetic materials as 
indeed observed in a number of measurements  Such a shortening 
was indeed observed in a number of  measurements (see 
Refs.~\cite{Kawaguchi,Soulen,Goldman,Lazar} and references 
therein).

Recently, new experiments \cite{Petrashov1,Petrashov2,Geroid}  
revealed  a large excess conductance of the F/S boundary, which 
was interpreted in terms of a long-range proximity effect 
in the ferromagnet. It was  pointed out \cite{Petrashov2} that 
spin triplet fluctuations in the electron-hole correlations caused by 
the spin-orbit interaction and electron-impurity  scattering 
\cite{Spivak} can  not (by two orders of magnitude) explain the 
large effect observed in Refs.\cite{Petrashov1,Petrashov2,Geroid}.
 
The main message of our paper is that in magnetically inhomogeneous 
materials (such as multi-domain ferromagnets, inhomogeneous 
"cryptoferromagnetic" states imposed by the superconductor
\cite{Larkin},  
F/S interfaces inducing  electronic  spin-flip processes  \cite{Zhu}
etc.), 
strong quantum fluctuations of the  electron and hole spins 
make the proximity effect less sensitive to the spin selection rule
that 
applies to 
 Andreev reflections.  As a result,  a strong long-range, spin-triplet 
proximity effect in  F/S structures persists on  a length scale
typical 
for $non-magnetic$ materials.
We estimate the conductance of such an F/S structure to be of the 
same order of magnitude 
   as  the conductance measured in experiments
\cite{Petrashov1,Petrashov2,Geroid}. Additional experiments with 
intentionally introduced magnetic inhomogeneities are needed to check 
the predicted effect quantitatively.
 
An analytical solution for the proximity effect was obtained for the case 
when spin scattering occurs at distances shorter than the electronic 
mean free path $l_0$. This allows us to consider perfect Andreev 
reflection and spin scattering using the Eilenberger equation and to
formulate 
proper boundary conditions for the complex scattering at magnetically 
inhomogeneous F/S interface. These boundary conditions were used to solve 
the Usadel equation in the diffusive region of the ferromagnet and for 
calculating the excess conductance of the F/S boundary. We show that a 
new type of superconducting ordering corresponding to  spin triplet 
correlations is the source of
 the proximity effect at  distances of order $L_T \gg L_{I_0}$ 
\footnote{One can show that the result expressed in terms of the 
spin-flip amplitude does not depend on the ratio between $l_0$ and 
$L_{I_0}$}. In order for an analytical solution to be feasible,  we 
have considered the case of weak magnetic scattering. This is not 
a  condition for the existance of the proposed long-range roximity 
effect,  but it allows us to conveniently  linearize the effective 
Usadel equation for the triplet components of the Green's function.

The Green's function of the problem, $ \hat{{\bf g}}$, is an $8\times 8$ 
matrix, which is the tensor product of the Keldysh $2 \times 2$
and the particle-hole (Nambu + spin) $4 \times 4$ matrices. Hence 
\begin{equation}
\hat{{\bf g}} =\left(\begin{array}{ll}
\hat{g}^R & \hat{g}^K \\
\hat{0} & \hat{g}^A
\end{array} \right) \, ,
\label{Keldysh}
\end{equation}
where $\hat{g}^{R,K,A}$ are  retarded (R), Keldysh (K) and advanced
(A) $4\times 4$ matrix Green's functions which include both the
singlet 
$\hat{g}_{\sigma ,-\sigma }$ and triplet $\hat{g}_{\sigma ,\sigma }$ 
components of  the normal as well as anomalous Green's functions. The 
pairing potential $\hat{{\bf \Delta}}$ determining electron-hole 
correlations, can be written in terms of Pauli matrices
$\hat{\sigma}_i $ 
and the superconducting energy gap
$\Delta$: $\hat{{\bf \Delta}}=\hat{\sigma}_0 \otimes 
\hat{\sigma}_3 \otimes \left(\Delta^{*} \hat{\sigma}_{-}  - 
\Delta \hat{\sigma}_{+}\right)$ with $\hat{\sigma}_{\pm} =
\hat{\sigma}_1 \pm \hat{\sigma}_2 $. The Eilenberger equation 
\cite{Eilenberger} in the ballistic region $x \leq b$ (see Fig.1) 
is written as 
\begin{equation}
i v_F {\bf n}\frac{ \partial }{\partial {\bf R} } \hat{{\bf g}}+
\left[\epsilon \hat{{\bf \tau}}_3 +\hat{{\bf \Delta}} -  \hat{{\bf
      h}}, 
\hat{{\bf g}} \right]= 0 \, ,
 \label{Eilen}
  \end{equation}
where $\epsilon$ is the energy measured from the Fermi level, ${\bf n}$  
is a unity vector along the electron momentum,  $\hat{{\bf h}}$ is the 
operator
that describes the effect of the inhomogeneous magnetic moment 
$\vec{h}(x)=I_0 \vec{e}(x)$ on the spins of electrons and holes 
($\vec{e}(x)$ is a unit vector): $\hat{{\bf h}} = \hat{\sigma}_0
\otimes 
\left(h_x \hat{\sigma}_1\otimes \hat{\sigma}_0 + h_y \hat{\sigma}_2 
\otimes \hat{\sigma}_0  + h_z\hat{\sigma}_3 \otimes \hat{\sigma}_3 
\right) $.

%%%%%%%%%%%%%%%%%%%%%%%
\begin{figure}
\centerline{\psfig{figure=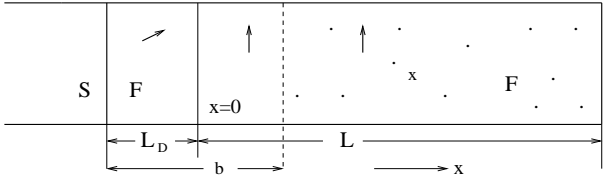,width=8cm}}
\vspace*{2mm}
\caption{Schematic view of an S/F structure with a domain wall
 at $x=0$, a distance $L_D$ from the S/F interface. Impurity
scattering is assumed to occur out of the ballistic region 
(to the right of the vertical dashed line).}
\end{figure}
%%%%%%%%%%%%%%%%%%%%%%%%

Under the assumption that "spin splitting" due to magnetic 
scattering  is small, one can find the solution of Eq.(\ref{Eilen}) 
in a ballistic region which includes a magnetic inhomogeneity
(Cf. Fig.1). 
Outside  the ballistic region, $x \gg l_0$, the Usadel equation
\cite{Usadel} 
controls the proximity effect  both for the singlet and the triplet 
components of the isotropic part of the matrix Green's function 
$\hat{{\bf G}}=\left<\hat{{\bf g}}\right>$ ($\left<...\right> $ 
denotes an average over the directions of electron/hole momenta). 
One finds  that while the singlet components decay at distances of 
order $L_{I_0}$, both   
the normal and anomalous spin triplet components persist at much 
longer distances of order $x \approx L_T$. Using this fact one can 
linearize the problem with respect to the triplet  components of the 
anomalous Green's function - the only ones that survive at distances  
much larger than $x \gg L_I{_0}$. As a result, one gets both the
Usadel equation and its boundary conditions \footnote{While deriving the 
boundary condition we used the fact that 
$\left<n_x \hat{{\bf g}}\right>$ does not depend on $x$ (\cite{Kuprianov}) 
and hence connects
solutions of the Eilenberger equation in the ballistic region $|x| \ll l_0$ 
with solutions of the Usadel equation in the diffusive region.} as
\begin{equation}
\hbar D \frac{d^2 \Theta_{\sigma}}{d x^2} -2i\epsilon \Theta_{\sigma } = 0
 \label{Usadel} 
  \end{equation}

 \begin{equation}
\frac{d}{d x} \Theta_{\sigma  }|_{x=0} =\sigma
\frac{v_F}{D}|\left<|n_x|r_{sf}\right>| \, .
\label{boundary} 
  \end{equation}
Here $r_{sf}$ is the magnetic spin-flip scattering  amplitude, 
$\sigma =\pm 1$. In the above equations we have used the standard
parametrization (see, {\em e.g.}, Ref. \cite{Belzig})
of the triplet Green's functions for normal pairing
$G_{\sigma,\sigma}^R $,  
$\bar{G}_{\sigma,\sigma}^R $ and  anomalous pairing   
$F_{\sigma,\sigma}^R $,  $\bar{F}_{\sigma,\sigma}^R $.  
It follows that
\begin{equation}
\hat{G}_{\sigma,\sigma}^R = \left(\begin{array}{cc}
\cosh (\Theta_{\sigma})  & \sinh (\Theta_{\sigma} )\exp(i \chi_{\sigma})\\
-\sinh (\Theta_{\sigma})\exp(-i \chi_{\sigma}) & -\cosh (\Theta_{\sigma})
\end{array}\right) \, ,
 \label{parametr} 
  \end{equation}
where $\Theta_{\sigma}$ and $\chi_{\sigma}$ are complex functions; the 
function $\chi_{\sigma}$  does not contribute to the conductance (see 
Eq.~(\ref{conductance}) below ). Equations (\ref{Usadel}) and 
(\ref{boundary}) are linear due to the 
smallness of the amplitude for magnetic spin-flip scattering,
$|r_{sf}| 
\ll 1$, and are valid in the temperature interval $k_BT \ll \Delta$
which includes the Thouless enegy $k_BT_{Th} = \hbar D/L^2 \ll \Delta$.

In order to calculate the spin-flip amplitude $r_{sf}$ one needs to
know 
the detailed character  of the magnetic inhomogeneity. A  quantitative 
theory  can  only be formulated in  case  the magnetic structure is
known 
in the experiment of interest.
In the absence of any precise information about the magnetic structure
of the samples  used in  existing experiments, 
we turn to illustrative examples of magnetic disorder and restrict 
ourselves to making  only  qualitative  comparisons with experiments. 
We will consider two such examples:
(i) the spin-splitting magnetic scattering is due to a  multi-domain 
structure with non-colinear magnetization in the neihgbouring domains  
and (ii)
 colinear magnetization of neighboring domains but with
 spin-splitting 
scattering in the domain wall.

 In case (i)  Rabi oscillations of the spin direction appear, and by  
solving  the Eilenberger equation one can show the probability
amplitude $r_{sp}$  to be
\begin{equation}
r_{sf} =(h^{(0)}_x +ih^{(0)}_y)/I_0) 
\sin\left(\frac{I_0 L_D}{\hbar v_F |n_x|}\right) \, .
 \label{Ruby} 
  \end{equation}
Here ${\bf h^{(0)}}$ is the magnetic moment in the domain closest to
the superconductor, $L_D$ is the width of the domain (see Fig.1).
Equation (\ref{Ruby}) is valid if the component of the electron/hole 
 momentum perpendicular to the interface  between the  domains is not
 too small:  $n_x^2 \gg I_0/\epsilon_F$; we  have neglected the
 contribution 
of the domain wall itself by appealing to the smallness of its width
$L_{dw}$,  $L_{dw} \ll L_D$.

 In case (ii) we have $h^{(0)}_x = h^{(0)}_y =0$, there are no Rabi 
oscillations (see Eq.~(\ref{Ruby})) and only the domain wall "split" 
the spin, provided that $L_{dw}$ is of the order or less than 
$\xi_{I_0} =\hbar v_F/I_0$. For the case   $L_{dw} \ll \hbar v_F
|n_x|/I_0$ the  probability
amplitude for  splitting caused by the domain wall  is 
\begin{equation}
r_{sf} =\frac{L_{dw} I_0}{\hbar v_F |n_x|} \, .
 \label{wall} 
  \end{equation}

In order to calculate the conductance we follow Ref.~\cite{Lambert}
and find that the excess  conductance can be written as 
\begin{equation}
\frac{\delta G}{G_N}= -\frac{1}{16 T}\sum_{\sigma=-1}^{1}
\int_{-\infty}^{\infty}
d\epsilon \frac{\partial f_0}{\partial \epsilon}\left(\frac{1}{L}
\int_{0}^L
dx (Re \Theta_{\sigma})^2 \right)\, .
 \label{conductance} 
  \end{equation}

The solution of the Usadel equation (\ref{Usadel})  with the
boundary condition given by Eq.(\ref{boundary}) at $x=0$ and by 
$\Theta_{\sigma} = 0$ at $x=L$, is

\begin{equation}
 \Theta_{\sigma  } =\sigma 
\frac{v_F |\left<|n_x| r_{sf}\right>|}{D k(\epsilon)}
\frac{\sinh (k(\epsilon)(x-L))}{\cosh (k(\epsilon)L)} \, ,
 \label{solution} 
  \end{equation}
where $k(\epsilon)= (1+i)\sqrt{\epsilon /\hbar D}$.

Equation (\ref{solution}) shows that the superconducting correlations 
due to the spin-splitting processes in the magnetic inhomogeneous
region 
decay exponentially in the ferromagnet and vanish at  distances of
order 
$L_T$ (for energies $\epsilon \sim kT$) corresponding to the 
 superconducting correlation length in non-magnetic materials.

Inserting Eq.~(\ref{solution}) into Eq.~(\ref{conductance}) one obtains 
an excess  conductance that can be expressed as
\begin{equation}
 \delta G / G_0 = \gamma f(T/T_{Th}) \, ,
\label{excess} 
  \end{equation}
where $$\gamma = |\left<|n_x|r_{sf}\right>|^2\left(L/l_0\right)^2 
$$
and   $f(T/T_{Th})$ is a dimensionless function, the  temperature 
dependence of which is presented in Fig.2,
\begin{equation}
  f(x)= \frac{1}{x} \int_0^{\infty} dt \cosh^{-2}(t^2/2x) 
\label{f} 
  \end{equation}
$$\times \left(Re\frac{\sinh(2(1+i)t)-2(1+i)t}{4(i-1)t^2\cosh^2([(1+i)t])}+\frac{\sinh 2t-\sin2t}{4t^2|\cosh(1+i)t|^2}\right) $$

 Using experimental values of the parameters taken from
 Ref.\cite{Petrashov1} $D=10$ cm$^2$/s and 
$T/T_{Th}=50$, and with the reasonable assumption that $r_{sf}\sim
 10^{-1} $ our result for 
the excess resistance $\delta R \approx -10$ Ohm is agreement with the 
experiment. The temperature dependence of the excess conductance in
 the  
range $T \sim T_{Th}$ is shown  in Fig.2. For higher temperatures,  
$T \sim \Delta/k_B \gg E_{Th}$ our theory is not valid and contributions
 of 
order
$kT/\Delta \sim 1$ can modify the temperature dependence of the
 resistance. Additional measurements around  the Thouless temperature 
(where the proximity effect is most pronounced) would permit a
 comparison  
with the temperature dependence coming from the long-range proximity 
effect described by our theory.    However, additional  investigations 
of the magnetic structure of the F/S interface  are needed to carry 
out a complete comparison with the theory. 
Multi-domain ferromagnets suitable for these studies can be created in 
various ways. It was recently demonstrated \cite{Zdravko} that grain 
boundaries, magnetic inhomogeneities
(including non-colinear magnetic domains) can be introduced in a 
predetermined position in a ferromagnet film by controlling the 
epitaxial growth.
Experiments, where  such magnetic inhomogeneities are intentionally 
created, would  permit long-range proximity effects to be studied 
 in ferromagnet-superconductor
structures.

%%%%%%%%%%%%%%%%%%%%%%%
\begin{figure}
\centerline{\psfig{figure=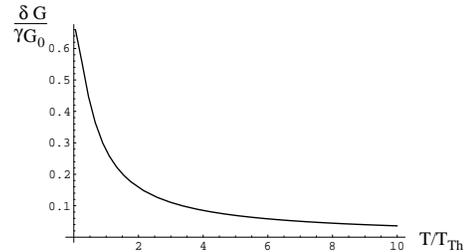,width=6cm}}
\vspace*{2mm}
\caption{Temperature dependence of the normalized excess conductance 
(see Eqs.(\ref{excess}) and (\ref{f})).}
\end{figure}
%%%%%%%%%%%%%%%%%%%%%%%%
%excess
In conclusion, we have shown  that  spin-splitting scattering related 
to magnetic inhomogeneities  modifies the sin-selection rule governing  
Andreev reflections at a ferromagnetic normal metal - superconductor interface.
As a result a long-range proximity effect, due to correlations between 
spin-aligned electrons and holes, 
appears (a spin triplet proximity effect).  Estimations of
the value of the excess conductance are consistent with experiments 
\cite{Petrashov1,Petrashov2,Geroid}.

After this work was completed we learned  that a similar problem have 
been  addressed recently \cite{Volkov}. Considering a somewhat
different model for magnetic inhomogeneity  and discussing  the case 
of a weak proximity effect (corresponding to a low-transparency tunnel 
barrier at the F/S interface) the authors of Ref.~\cite{Volkov} 
have come to the same conclusion 
as we have when it comes to the existence of a long-range triplet 
proximity effect in magnetically inhomogeneous ferromagnets. 

We acknowledge useful discussions with  E.V. Bezuglyi and
Z.G. Ivanov. 
We are grateful to Z.G. Ivanov for  calling to our attention  
experimental possibilites
to observe the effect predicted by our theory.

\end{document}